\documentclass[twocolumn,tighten,times]{aastex631}

\usepackage[running,mathlines]{lineno}

\usepackage{lipsum}

\shorttitle{Afterglow QPO filtering}
\shortauthors{Globus}

\graphicspath{{./}{figures/}}

\begin{document}

\title{Fireballs’ Whispers of Their Central Engine: Relativistic Filtering of Afterglow QPOs}

\author[0000-0001-9011-0737]{No\'emie Globus}
\affiliation{Instituto de Astronom{\'\i}a, Universidad Nacional Aut\'onoma de M\'exico, km 107 Carretera Tijuana-Ensenada, 22860, Ensenada, M\'exico}

\begin{abstract}
Quasi-periodic oscillations (QPOs) in gamma-ray bursts (GRBs) afterglows  have been suggested as probes of the central engine. Such interpretations generally assume that the observed modulation frequency directly corresponds to an intrinsic oscillation frequency of the
source. We show that this assumption is not generally valid and that interpreting such features without accounting for relativistic propagation may  lead to misleading inferences about the engine nature. We show that relativistic propagation effects—most importantly integration over
equal-arrival-time surfaces—act as a frequency-dependent filter that can
significantly modify or suppress intrinsic variability. In the constant-$\Gamma$ case, the angular kernel acts as a stationary low-pass filter that suppresses high-frequency variability without altering its frequency, whereas Blandford–McKee deceleration renders the filter time-dependent and manifests observationally as an apparent frequency drift. 
\end{abstract}


\section{Introduction}

Following the prompt phase of a gamma-ray burst (GRB), the afterglow arises from the interaction of a relativistic jet with the external medium, producing broadband synchrotron emission from X-ray to radio wavelengths \citep[][for a review]{2004RvMP...76.1143P}. Afterglow light curves are typically smooth and decay at early times, sometimes exhibiting plateaus. Rarely, quasi-periodic oscillations (QPOs) have been reported at early times \citep{2025ApJ...985...33G,2025MNRAS.541.3787S}, and are often interpreted as directly tracing intrinsic oscillations of the  central engine.

However, relativistic beaming and photon arrival-time effects
\citep[e.g.,][]{LindBlandford1985, Fenimore1996, Sari1998, PanaitescuKumar2000} inevitably smooth temporal structure in relativistic outflows, strongly attenuating and potentially distorting high-frequency variability. Consequently, the presence, amplitude, and even the characteristic frequency of a quasi-periodic modulation observed in the afterglow do not in general correspond directly to those of the underlying central-engine oscillation. Despite their central role in afterglow modeling, these relativistic propagation effects have not been explicitly formulated in the context of QPOs, nor cast as a temporal transfer function mapping intrinsic variability to the observed signal.

In this \textit{Letter}, we derive a temporal transfer function framework, linking intrinsic variability to the observed afterglow signal. We show that relativistic propagation alone can generate apparent frequency drift in afterglow QPOs, even when the engine oscillation frequency is constant. As a result, observed QPO periods cannot, in general, be directly mapped to intrinsic engine timescales.

The \textit{Letter} is organized as follows. In Section~\ref{sec:eats}, we formulate a linear response kernel based on equal–arrival–time surfaces that maps intrinsic central-engine variability to the observed afterglow. We apply this framework to a constant–Lorentz-factor flow, appropriate for the coasting (pre-deceleration) phase, in Section~\ref{sec:coasting}. In Section~\ref{sec:bm}, we show that the \citet{Blandford1976} (hereafter, BM) deceleration renders the filter time-dependent, leading to a systematic drift in both the modulation amplitude and the observed frequency. Section~\ref{sec:offaxis} generalizes the analysis to non-uniform shell emission. We discuss a specific GRB in Section~\ref{220711B} before concluding in Section~\ref{Conclusion}.

\section{Transfer function formulation}\label{sec:eats}
The observed emission from a GRB afterglow at cosmological redshift $z$ arises from a relativistically
expanding blast wave. Due to light--travel--time effects and relativistic beaming,
photons emitted simultaneously in the laboratory frame from different locations
on the blast--wave surface arrive at the observer at different times. As a result,
radiation received at a fixed observer time originates from an
\emph{equal--arrival--time surface} (EATS; \citealt{Panaitescu1998, SariPiran1997, Sari1998}). The EATS is defined by the photon arrival--time condition
\begin{equation}
t_{\rm obs}
=
(1+z)\left(
t'
+
\frac{D - \mathbf{n}\!\cdot\!\mathbf{x}}{c}
\right),
\label{eq:arrival_condition}
\end{equation}
where $t'$ is the emission time in the lab frame, $D$ is the distance from the
explosion center to the observer,
$\mathbf{x}$ is the emission position ($|\mathbf{x}| \ll D$), and $\mathbf{n}$ is the unit vector toward
the observer. The observed specific flux at observer time $t_{\rm obs}$ may be written as an
integral over the EATS,
\begin{equation}
F_{\nu}(t_{\rm obs}) =
\frac{1}{4\pi D_L^2 (1+z)}
\int_{\rm EATS} dV'\,
j'_{\nu'}\,\mathcal{D}^3 ,
\label{eq:Fnu_exacteats}
\end{equation}
\citep[e.g.,][]{GranotPiranSari1999,PanaitescuKumar2000},
where $D_L$ is the luminosity distance, $j'_{\nu'}$ is the comoving emissivity at
$\nu'=\nu(1+z)/\mathcal{D}$, $\mathcal{D}=[\Gamma(1-\beta\cos\theta)]^{-1}$ is the Doppler
factor.  

Let us introduce a new arrival-time variable, $T\equiv t_{\rm obs}/(1+z)-D/c$, which absorbs the constant $D/c$ and the redshift correction; Eq.~\ref{eq:arrival_condition} then becomes 
\begin{equation}
T = t' - \frac{\mathbf{n}\!\cdot\!\mathbf{x}}{c}.
\label{eq:T_def}
\end{equation}

The integral in Eq.~\ref{eq:Fnu_exacteats} over the EATS may equivalently be written as an integral over all
spacetime with a delta function enforcing the arrival--time constraint,
\begin{eqnarray}
F_{\nu}(T)
&&=
\int dt' \int d^3x\;
\frac{\mathcal{D}^3}{4\pi D_L^2 }
\delta\!\left(
T-
t'
+ \frac{\mathbf{n}\!\cdot\!\mathbf{x}}{c}
\right)
j'_{\nu'}(\mathbf{x}, t')\nonumber\\
&&\equiv \int dt'
\int d^{3}x \;
G_{\nu}(T; \mathbf{x}, t') \,
j'_{\nu'}(\mathbf{x}, t') ,
\label{eq:Fnu_greeneats}
\end{eqnarray}
where the Green's function $G_{\nu}(T; \mathbf{x}, t')$
encodes the light--travel--time delays, angular dependence, and relativistic
Doppler boosting.
It is useful to isolate how small time-dependent
perturbations at the engine propagate into the observed light curve. We assume
that the comoving emissivity can be written as a small perturbation about a
steady background,
\begin{equation}
j'_{\nu'}(\mathbf{x},t')
=
\bar{j}'_{\nu'}(\mathbf{x})
\left[1+\epsilon\,S(t')\right],
\qquad
\epsilon \ll 1 ,
\end{equation}
where $S(t')$ describes the engine-frame modulation. When the perturbation
amplitude is small and the Lorentz factor varies slowly across a single
equal–arrival–time surface, the response is linear.
Linearizing  Eq.~\ref{eq:Fnu_greeneats} in 
$\epsilon$ yields,
\begin{equation}
\delta F(T)
=
\int dt'\,
S(t')
\int d^{3}x\;
G(T;\mathbf{x},t')\,
\bar{j}'_{\nu'}(\mathbf{x}) .
\end{equation}
This defines an effective response kernel,
\begin{equation}
G(T,t')
\equiv
\int d^{3}x\;
G_{\nu}(T;\mathbf{x},t')\,
\bar{j}'_{\nu'}(\mathbf{x}) ,
\end{equation}
which acts as a transfer function relating engine-frame variability to the
observed light curve.

\section{Coasting GRB blast waves}\label{sec:coasting}

We adopt the standard ultra-relativistic, thin-shell approximation appropriate for GRB afterglows, assuming spherical symmetry, 
\(\Gamma \gg 1\), \(\theta \ll 1\), and treating the Lorentz factor as approximately constant across a single EATS. In the thin-shell limit, the photon arrival time from a radius \(R\) and angle \(\theta\) is
\begin{equation}
T \simeq \frac{R}{2c\Gamma^2} + \frac{R\theta^2}{2c} \equiv T^{(0)} + \tau(\theta),
\label{eq:T_ultra1}
\end{equation}
where \(T^{(0)}\) is the on-axis arrival time and \(\tau(\theta)\) represents the angular light-travel-time delay  associated with the curvature of the emitting surface. Radiation emitted at angle \(\theta\) is Doppler boosted by 
\begin{equation}
\mathcal{D}(\theta) \simeq \frac{2\Gamma}{1 + \Gamma^2 \theta^2},
\end{equation}
which suppresses contributions from large angles.
This decomposition of observed arrival time into radial and angular light‑travel delays follows standard treatments of relativistic blast waves \citep[e.g.,][]{2004RvMP...76.1143P}.

Assuming a thin radial shell, the spatial integral reduces to an integral over solid angle $\delta\Omega$. 
Using spherical coordinates
$d^3x = r^2 dr\, d\Omega$ and inserting the radial delta function yields
\begin{eqnarray}
\int d^3x\,\bar{j}'_{\nu'}(\mathbf{x})(\cdots)
&&=
\int r^2 dr\, d\Omega\;
\delta(r-R)\,\bar{j}'_{\nu'}(R)(\cdots)\nonumber\\
&&=
R^2 \bar{j}'_{\nu'}(R)\int d\Omega\,(\cdots).
\end{eqnarray}

The response kernel therefore becomes
\begin{equation}
G(T,t')
\propto
\int d\Omega\;
\mathcal{D}^3(\theta)\,
\delta\!\left[
T - t' + \frac{R\cos\theta}{c}
\right].
\end{equation}
Using the ultra--relativistic expression for the arrival time
(Eq.~\ref{eq:T_ultra1}), we write
\begin{equation}
T = T^{(0)} + \tau(\theta),
\qquad
\tau(\theta) \equiv \frac{R}{2c}\theta^2,
\end{equation}
so that the kernel depends only on the angular delay $\tau$. The delta function
may then be rewritten as $\delta\!\left[T - T^{(0)} - \tau(\theta)\right]$. Since we assumed small angles ($\theta \ll 1$), the solid-angle element becomes
\begin{equation}
d\Omega = 2\pi \sin\theta\, d\theta \simeq 2\pi \theta\, d\theta
= 2\pi \frac{c}{R}\, d\tau,
\end{equation}
where we have used $\tau = R\theta^2/(2c)$. The response kernel therefore reduces
to a one-dimensional integral over angular delay,
\begin{equation}
G(T)
\propto
\int_0^\infty d\tau\;
\mathcal{D}^3(\tau)\,
\delta\!\left[T  - T^{(0)} - \tau\right].
\end{equation}

Using $\mathcal{D}(\theta) = 2\Gamma/(1+\Gamma^2\theta^2)$ and
$\Gamma^2\theta^2 = \tau/\tau_0$ with
\begin{equation}
\tau_0 \equiv \frac{R}{2c\Gamma^2},
\label{tau0}
\end{equation}
the Doppler factor becomes
\begin{equation}
\mathcal{D}(\tau) = \frac{2\Gamma}{1+\tau/\tau_0}.
\end{equation}
Up to an overall normalization, the angular response kernel therefore takes the
form
\begin{equation}
K(\tau) \propto \left(1+\frac{\tau}{\tau_0}\right)^{-3}.
\end{equation}

Imposing the normalization condition
$\int_0^\infty K(\tau)\,d\tau = 1$ yields the fully normalized angular-delay
kernel
\begin{equation}\label{kernel}
K(\tau)
=
\frac{2}{\tau_0}
\left(1+\frac{\tau}{\tau_0}\right)^{-3}.
\end{equation}

Therefore, in the thin-shell, constant-$\Gamma$ limit, the observed variability is thus a
causal convolution of the intrinsic engine-frame signal with this angular-delay
kernel,
\begin{equation}
F(T)
=
\int_0^\infty d\tau\;
K(\tau)\,
S\!\left(T - T^{(0)} - \tau\right).
\end{equation}

 The kernel $K(\tau)$ may be interpreted as the angular-delay slice of the full response function $G$, obtained by integrating over the EATS at fixed arrival-time delay. It describes how  relativistic beaming and light--travel--time effects smooth rapid intrinsic variability.

In the coasting phase considered here, the bulk Lorentz factor is constant and
the angular response kernel is stationary, depending only on the angular
light--travel--time delay $\tau$ (Eq.~\ref{kernel}). Under this assumption, it is
meaningful to characterize the response in the Fourier domain. Time-dependent
effects arising during Blandford--McKee deceleration, which render the kernel
non-stationary, are treated separately in Section~\ref{sec:bm}.

The Fourier
transform of the kernel defines the transfer function (Appendix~\ref{appendixA}), which reduces at low frequencies to
\begin{equation}
H(\omega)\equiv
\frac{\widetilde{K}(\omega)}{\widetilde{K}(0)}
\simeq \frac{1}{1+i\omega\tau_0}.
\label{H_omega}
\end{equation}

The dimensionless function $H(\omega)$ is the transfer function of the system.
It quantifies how sinusoidal variability at angular frequency $\omega$ is filtered
by light-travel-time effects. 
The low-frequency form corresponds to a first-order low-pass filter with cutoff frequency $\omega\sim\tau_0^{-1}$.

\section{Perturbative inclusion of Blandford--McKee deceleration}\label{sec:bm}

In the preceding sections we assumed that the bulk Lorentz factor is approximately
constant across a single EATS. We now relax this assumption and incorporate the
effects of gradual blast--wave deceleration. When the intrinsic variability
timescale of the central engine is short compared to the deceleration timescale,
these effects may be treated perturbatively.

For an adiabatic BM blast wave propagating into a uniform external
medium, the Lorentz factor and shock radius evolve with observer time as \citep{Blandford1976}:
\begin{equation}
\Gamma(T) \propto T^{-3/8},
\qquad
R(T) \propto T^{1/4}.
\end{equation}

The characteristic angular delay—defined as the arrival--time difference between
photons emitted on--axis and those emitted at angles
$\theta \sim \Gamma^{-1}$—is
\begin{equation}
\tau_0(T)
\equiv
\frac{R(T)}{2\Gamma^2(T)c}
\propto T .
\end{equation}
Thus, in a self--similar decelerating flow, the angular light--travel--time scale
grows linearly with observer time.

Because both $\Gamma$ and $R$ evolve with time, the angular response kernel is no
longer strictly invariant under time translations. However, over intervals short
compared to the deceleration timescale, the blast wave responds locally as if it
were characterized by a constant Lorentz factor, with a slowly varying angular
delay scale $\tau_0(T)$. This separation of timescales justifies a
quasi--stationary (adiabatic) approximation for the response kernel.

To illustrate the resulting observational consequences, consider a small
fractional modulation of the comoving emissivity driven by the central engine,
\begin{equation}
\frac{\delta S}{S}
=
\varepsilon \sin(\omega t'),
\qquad
\varepsilon \ll 1 .
\end{equation}
To leading order in $\varepsilon$, the observed fractional flux variation is
obtained by convolving this modulation with the angular--delay kernel evaluated at
the instantaneous value of $\tau_0$. The resulting response is
\begin{equation}
\frac{\delta F}{F}(T)
\simeq
\varepsilon
\left[1 + (\omega \tau_0)^2 \right]^{-1/2}
\sin\!\left[
\omega t'(T) - \tan^{-1}(\omega \tau_0)
\right],
\label{eq:flux_response_BM}
\end{equation}
which exhibits both amplitude suppression and a frequency--dependent phase lag relative to the engine intrinsic modulation. As a result, the observed instantaneous
    frequency,
    \[
    \omega_{\rm obs}(T) =
    \frac{d}{dT}
    \Big[
    \omega\, t'(T)
    -
    \tan^{-1}\big(\omega \tau_0(T)\big)
    \Big],
    \]
    decreases slowly with arrival time relative to the engine frequency,
    reflecting the combined effects of blast--wave deceleration and angular
    light--travel--time delays.

For a decelerating BM blast wave, both the Lorentz factor and the
angular--delay timescale evolve with time. The observed phase is now
\begin{equation}
\phi(T)=\omega\,t'(T)-\tan^{-1}\!\big[\omega\tau_0(T)\big],
\end{equation}
where $t'$ is the laboratory--frame emission time. Differentiating with respect to
the observer time $T$ gives
\begin{equation}
\omega_{\rm obs}(T)
=
\omega\,\frac{dt'}{dT}
-
\frac{\omega\,\dot{\tau}_0(T)}{1+\big(\omega\tau_0\big)^2},
\label{eq:omega_obs_BM}
\end{equation}
Unlike the constant--$\Gamma$ case, the observed frequency now decreases secularly with time, implying a gradual increase of the observed period driven
by both blast--wave deceleration and the growth of the angular--delay timescale.

\begin{figure}
\centering
\includegraphics[width=0.5\textwidth]{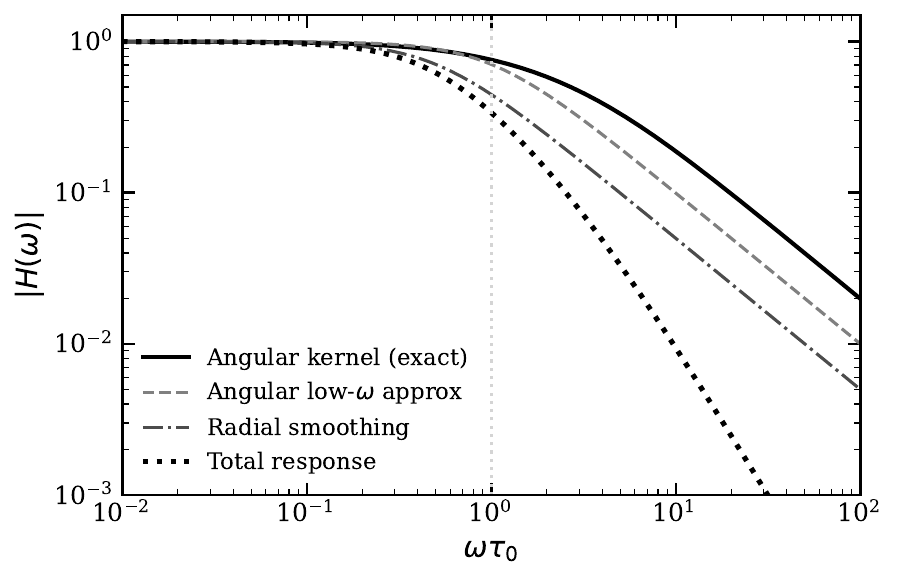}
\caption{Magnitude of the transfer function $|H(\omega)|$ for the angular-delay kernel. 
In Fourier space, the observed variability is obtained by multiplying the Fourier transform of the central-engine signal by $|H(\omega)|$, which reduces the amplitude of each frequency component while leaving the frequency itself unchanged in the constant-$\Gamma$ coasting phase. 
The solid black line shows the exact angular kernel, the dashed gray line shows the low-frequency approximation $|H|\simeq1/\sqrt{1+(\omega \tau_0)^2}$, and the vertical dotted gray line marks $\omega \tau_0 = 1$, the approximate cutoff where high-frequency variability is suppressed. 
The dash--dotted curve illustrates the additional radial smoothing, and the dotted curve shows the combined response when radial and angular smoothing are both present. In this case the suppression begins slightly earlier, with the response declining for frequencies $\omega\tau_0 \sim 0.5$--1. Each causal smoothing process suppresses high-frequency variability as $|H|\propto \omega^{-1}$, so the combination of radial and angular delays produces an asymptotic suppression $|H|\propto \omega^{-2}$.}

\label{transferfunction}
\end{figure}

Radial smoothing, which arises from the finite width of the flow and becomes increasingly important as the ejecta spreads, is an additional effect that should be considered when interpreting variability. 
Hydrodynamical studies of relativistic GRB ejecta show that internal velocity differences cause the shell to begin spreading at a radius $R \sim \Gamma^{2} c t_{\rm engine}$, where $t_{\rm engine}$ is the duration of the central engine activity. Its width then grows with radius as $\Delta R \sim R/\Gamma^{2}$ \citep[e.g.,][]{Kobayashi1999}, leading to a radial light-travel-time delay comparable to angular curvature effects. In our formalism, such an additional smoothing can be approximated, to first order, as a second transfer function with a characteristic width comparable to that of the angular curvature kernel. In this simplified limit, the total response can be approximated as the product of the angular and radial transfer functions. Because the associated timescales are of the same order, the characteristic cutoff frequency remains of order $\omega \sim c\Gamma^2/R$, but the combined filtering produces a steeper suppression of variability at high frequencies than curvature effects alone. This behavior is illustrated in Fig.~\ref{transferfunction}.  While a complete treatment of radial smoothing is beyond the scope of this paper, as it would require a full hydrodynamic treatment of interacting shells, this simple estimate indicates that radial effects cannot be neglected and are expected to contribute significantly to the filtering of QPOs.

\section{Emission from a Small Angle: Off-Axis Observers or Patchy Shells}\label{sec:offaxis}

Under the standard assumption of uniform emissivity over the full EATS, the dominant contribution to the observed flux extends out to angles $\theta \sim \Gamma^{-1}$, yielding the familiar angular-delay timescale $\tau_0 = R/(2c\Gamma^2)$. If the emission is instead confined to a small angular region 
$\theta_{\rm eff} \ll \Gamma^{-1}$ within the relativistic beaming cone,
as occurs for angularly localized (“patchy”) emission or slightly misaligned
emitting regions, the relevant light-travel-time spread is reduced to
\begin{equation}
\tau_{\rm eff} \simeq \frac{R}{2c}\,\theta_{\rm eff}^2 \ll \tau_0 \, .
\end{equation}

Here, $\theta_{\rm eff}$ represents the angular extent of the emitting region that effectively contributes to the observed flux.

\section{The case of  GRB~220711B}\label{220711B}

 As an illustrative example, consider the 50\,s QPO reported in the X-ray afterglow of GRB~220711B, which motivated this study. \citet{2025ApJ...985...33G} find that the QPO are strong from $T_0 + 93~{\rm s}$ to $T_0 + 270~{\rm s}$ with a decreasing amplitude between 93s and 270s, and that a high-to-low frequency drift appears after 270~s. This would be marking the transition between the coasting and BM deceleration phases in our model. For typical external-shock parameters in a uniform medium with density $n \sim 0.1$--$1\,{\rm cm^{-3}}$, a deceleration time of a few hundred seconds corresponds to an initial Lorentz factor $\Gamma_0 \sim 10^2$ and a shock radius $R \sim 10^{16}$--$10^{17}\,\mathrm{cm}$ at the onset of the BM phase, values broadly consistent with those commonly inferred for \emph{Swift}/XRT afterglows.
 
 The GRB redshift is unknown, making it difficult to infer the intrinsic period and make a complete analysis, but we can give an illustrative example. Assuming $z = 2$, the engine-frame period is $P_{\rm eng} \simeq 16.7~{\rm s}$ and the amplitude ratio would then be $|H(T_2)|/|H(T_1)| \simeq 0.35$. The predicted factor-of-three amplitude decrease from $\sim 90$~s to $\sim 270$~s aligns qualitatively with the \emph{Swift}/XRT light curve, where QPO peaks fade while frequency remains nearly constant, as expected from relativistic angular smearing in the coasting regime. At $T \gtrsim 300$~s, the oscillations weaken further and drift to lower frequencies, signaling blast-wave deceleration. While a detailed  analysis is beyond this work and would be difficult without knowledge of the redshift, the QPO evolution is naturally explained by our relativistic filtering framework.

\section{Conclusion}\label{Conclusion}

We have presented a linear--response framework to quantify how relativistic
propagation filters intrinsic variability in GRB afterglows. By expressing the
observed flux as a convolution over EATSs, we derived an exact angular
response kernel and its Fourier transform.  
During BM deceleration, the kernel evolves in time, producing a secular drift of
the observed period and frequency. These features motivate a more detailed
analysis, which our treatment captures in several ways:
\begin{itemize}
    \item \textit{Exact mapping of engine-frame variability to observer time:} 
    Instead of a single timescale, we provide the full transfer function $H(\omega)$ (Appendix~\ref{appendixA}), which quantitatively describes how each frequency component of the intrinsic engine signal is suppressed. This allows one to compute the attenuation factor for arbitrary engine modulation periods, not just infer whether $\omega \tau_0 \gtrsim 1$.

    \item \textit{Time-dependent response during deceleration:} 
    For a BM blast wave, $\tau_0(T)$ grows with observer time as $\tau_0 \propto T$, producing a secular drift of the observed frequency (Eq.~\ref{eq:omega_obs_BM}). The standard curvature-timescale argument provides only a static estimate and does not capture this slow evolution.
\end{itemize}

The present treatment assumes that the engine modulates the total energy of the blast wave by a small fractional amount ($\epsilon \ll 1$), perturbing the comoving emissivity without significantly altering the global dynamics. In this limit, the Lorentz factor of the blast wave deviates only at order $\epsilon$ from the standard BM solution, so the EATS geometry is only weakly modified. 
When the engine remains dynamically coupled to the blast wave—as in the commonly discussed refreshed-shock model—slower ejecta released by the engine can catch up with the decelerating blast wave and inject additional energy \citep{Rees1998,Sari2000}. Hydrodynamic communication across the shocked region occurs on a comoving sound-crossing time, which corresponds to an observer-frame delay of order  $R/(c\Gamma^2)$, comparable to the curvature  timescale. These processes therefore smooth variations on a similar scale, implying that variability with $\omega \tau_0 \gtrsim 1$ is strongly suppressed.

If the energy injection becomes sufficiently strong, however, the blast-wave dynamics may depart significantly from the BM solution, and the EATS geometry can be substantially distorted. The present framework is not intended to describe highly non-linear variability associated with strong refreshed shocks, large-amplitude energy injection, or sharp external-density discontinuities. Likewise, strongly structured jets may introduce additional angular weighting that modifies the detailed kernel shape. Nevertheless, whenever the variability amplitude remains modest and the dynamical evolution is smooth on the angular light-travel-time scale, the geometric filtering described here provides a robust description of how intrinsic engine variability maps onto the observed afterglow signal.

While detailed implications for central-engine models depend on uncertain parameters and will be addressed elsewhere, this \textit{Letter} establishes the relativistic filtering framework and assesses when QPOs—the faint whispers of the central engine—can survive propagation through the fireball and remain observable.

\bigskip
\bigskip
\section*{Acknowledgements}
We thank Roger Blandford for his insightful comments and for encouraging me to present this calculation in a formal publication. We thank the referee for a thoughtful report that has significantly improved the clarity of the manuscript.
We gratefully acknowledge the support of the Simons Foundation (MP-SCMPS-00001470, N.G.).

\bibliographystyle{aasjournal}
\bibliography{references}

\appendix
\section{Fourier transform of the angular-delay kernel}\label{appendixA}

We consider the causal delay kernel
\begin{equation}
K(\tau)=K_0\left(1+\frac{\tau}{\tau_0}\right)^{-3},
\qquad \tau\ge 0 ,
\end{equation}
with Fourier transform defined as
\begin{equation}
\widetilde{K}(\omega)
\equiv
\int_0^\infty d\tau\; K(\tau)\,e^{i\omega\tau}.
\end{equation}
Substituting for $K(\tau)$ gives
\begin{equation}
\widetilde{K}(\omega)
=
K_0 \int_0^\infty
\left(1+\frac{\tau}{\tau_0}\right)^{-3}
e^{i\omega\tau}\,d\tau .
\end{equation}

Introducing the change of variables
\begin{equation}
u = 1 + \frac{\tau}{\tau_0},
\qquad
\tau=\tau_0(u-1),
\qquad
d\tau=\tau_0\,du ,
\end{equation}
the integral becomes
\begin{equation}
\widetilde{K}(\omega)
=
K_0\,\tau_0\,e^{-i\omega\tau_0}
\int_1^\infty
u^{-3}\,e^{i\omega\tau_0 u}\,du .
\end{equation}

Using the definition of the upper incomplete gamma function,
\begin{equation}
\Gamma(s,z)=\int_z^\infty t^{s-1}e^{-t}\,dt ,
\end{equation}
and the substitution $t=-i\omega\tau_0 u$, one finds
\begin{equation}
\int_1^\infty u^{-3} e^{i\omega\tau_0 u}\,du
=
(\omega\tau_0)^2
e^{i\omega\tau_0}
\Gamma(-2,i\omega\tau_0).
\end{equation}
The Fourier transform therefore evaluates to
\begin{equation}
\widetilde{K}(\omega)
=
K_0\,\tau_0(\omega\tau_0)^2
e^{i\omega\tau_0}
\Gamma(-2,i\omega\tau_0).
\end{equation}

At zero frequency,
\begin{equation}
\widetilde{K}(0)
=
\int_0^\infty K(\tau)\,d\tau
=
\frac{K_0\tau_0}{2}.
\end{equation}
Defining the normalized transfer function
\begin{equation}
H(\omega)\equiv
\frac{\widetilde{K}(\omega)}{\widetilde{K}(0)},
\end{equation}
and using the small-argument expansion
\begin{equation}
\Gamma(-2,z)
\simeq
\frac{1}{2}z^{-2}-\frac{1}{z}+\mathcal{O}(1),
\qquad |z|\ll1 ,
\end{equation}
one obtains, for $\omega\tau_0\ll1$,
\begin{equation}
H(\omega)\simeq \frac{1}{1+i\omega\tau_0}.
\end{equation}

\end{document}